\documentclass[pra,aps,twocolumn,amsmath,amssymb,superscriptaddress,showpacs]{revtex4-1}
\usepackage{graphicx}
\usepackage{color}
\usepackage{textcomp}
\usepackage{bm}
\usepackage{hyperref}

\begin{document}

%%%%%%%%%%%%%%%%%%%%%%%%%%%%%%%%%%%%%% TITLE   %%%%%%%%%%%%%%%%%%%%%%%%%
\title{Superfluid--Insulator Transition in Strongly Disordered One-dimensional  Systems}
%%%%%%%%%%%%%%%%%%%%%%%%%%%%%%%%%%%%%% TITLE   %%%%%%%%%%%%%%%%%%%%%%%%%

%%%%%%%%%%%%%%%%%%%%%%%%%%%%%%%%%%%%%% AUTHORS %%%%%%%%%%%%%%%%%%%%%%%%%
\author{Zhiyuan Yao}
\affiliation{Department of Physics, University of Massachusetts, Amherst, MA 01003, USA} 

\author{Lode Pollet}
\affiliation{Department of Physics, Arnold Sommerfeld Center for Theoretical Physics and Center for NanoScience, University of Munich, Theresienstrasse 37, 80333 Munich, Germany}

\author{Nikolay Prokof'ev}
\affiliation{Department of Physics, University of Massachusetts, Amherst, MA 01003, USA} 
\affiliation{Russian Research Center ``Kurchatov Institute,'' 123182 Moscow, Russia}

\author{Boris Svistunov}
\affiliation{Department of Physics, University of Massachusetts, Amherst, MA 01003, USA}
\affiliation{Russian Research Center ``Kurchatov Institute,'' 123182 Moscow, Russia}
\affiliation{Wilczek Quantum Center, Zhejiang University of Technology, Hangzhou 310014, China}

%%%%%%%%%%%%%%%%%%%%%%%%%%%%%%%%%%%%%% AUTHORS %%%%%%%%%%%%%%%%%%%%%%%%%

\begin{abstract}
We present an asymptotically exact renormalization-group theory of the superfluid--insulator transition in one-dimensional disordered systems, with emphasis on an accurate description of the interplay between the Giamarchi--Schulz (instanton--anti-instanton) and weak-link (scratched-XY) criticalities. Combining the theory with extensive quantum Monte Carlo simulations allows us to  shed new light on the ground-state phase diagram of the one-dimensional disordered Bose-Hubbard model at unit filling.
\end{abstract}

  \maketitle
% 64.70.Tg Quantum phase transitions (for quantum Hall effects aspects, see 73.43.Nq in electronic structure of surfaces, interfaces, thin films, and low dimensional structures)

% 67.85.-d	Ultracold gases, trapped gases (see also 03.75.-b Matter waves in quantum mechanics)

% 67.85.Hj Bose-Einstein condensates in optical potentials

%%%%%%%%%%%%%%%%%%%%%%%%%%%%%%%%%%%%%%%

% 03.75.Hh Static properties of condensates; thermodynamical, statistical, and structural properties

% 05.30.Jp Boson systems (for static and dynamic properties of Bose-Einstein condensates, see 03.75.Hh and 03.75.Kk; see also 67.10.Ba Boson degeneracy in quantum fluids)

%%%%%%%%%%%%%%%%%%%%%%%%%%%%%%%%%%%%%%%%%%%%%%%%%%%%%%%%%%%%%%%%%%%%%
\section{Introduction}
\label{sec:I}

In 1960, Girardeau established  that in one dimension (1D) there is no qualitative difference between fermions and bosons:
spinless fermions can be exactly mapped onto hard-core bosons \cite{Girardeau}. Two decades later, Haldane demonstrated that
the low-energy physics of 1D superfluids is accurately captured by the Luttinger liquid (LL) paradigm, playing the role akin to that of the Fermi liquid paradigm in higher dimensions \cite{Haldane}. The universal character of the LL description of 1D superfluids becomes especially transparent after identifying the fermion-specific notion of backscattering events with quantum phase slippages (or instantons), and, correspondingly, associating the difference between the right- and left-moving fermions (in a system with periodic boundary conditions) with the winding number of the superfluid phase.
In this way, the LL picture reduces to quantized hydrodynamics augmented with phase slippages.  At the macroscopic level, it is via the instanton--anti-instanton pairs that superfluid hydrodynamics is coupled to either a commensurate external potential, or disorder, or both  (see, {\it e.g.}, \cite{Kashurnikov96,book2015}). For a LL in the infinite-size limit, the coupling renormalizes to zero in view of the absence of infinitely large [in the sense of the $(1+1)$-dimensional mapping] instanton--anti-instanton pairs. On approach to the critical point of a superfluid--insulator transition, large instanton--anti-instanton pairs become progressively more important until infinitely large pairs dissociate, causing the transition. Within Popov's $(1+1)$-dimensional hydrodynamic action over the phase field $\Phi(x, \tau)$, the instantons appear as  vortices (with a specific  $x$-dependent phase), such that
superfluid--insulator transitions are identical (in case of a Mott transition in a pure, commensurate system), or very close [in case of a superfluid(SF)--to--Bose-glass(BG) transition in a generic disordered system] to the Berezinskii--Kosterlitz--Thouless (BKT) transition. In analogy to the universal Nelson--Kosterlitz relation at the BKT point,
each type of the instanton--anti-instanton dissociation transition is characterized by a universal critical value $K_c$ of the Luttinger liquid parameter $K=\pi \sqrt{\Lambda \kappa}$ (with $\Lambda$ the superfluid stiffness  and $\kappa$ the compressibility). For the Mott transition in a pure system with commensurate filling $q/p$ ($q$ and $p$ are co-prime integers), this  universal value is $K_c=2/p^2$ \cite{Col} whereas  for the SF-BG transition it is $K_c=3/2$ \cite{GS}. Another well known and relevant phenomenon that is perfectly understood from the renormalization group (RG) analysis is
the Kane--Fisher renormalization of a single weak link (or an impurity) \cite{KaneFisher}: Below the universal value
$K_c^{\rm KF}=1$, a single arbitrarily weak impurity gets renormalized to an infinitely high (in {\it relative} low-energy units) barrier.
For $K > K_c^{\rm KF}$, by contrast, the link gets progressively healed with increasing length scale and becomes asymptotically transparent.

A controlled theory of SF-BG transition in 1D, yielding, in particular, $K_c=3/2$, was first developed by Giamarchi and Schulz (GS) using a perturbative RG treatment of disorder \cite{GS}. In the same work, the authors conjectured that there might exist an alternative strong-disorder scenario not captured by their theory. Subsequently, some of us demonstrated \cite{Kashurnikov96} that the GS result is valid beyond the lowest-order RG equations and is, in fact, a
generic answer thanks to the above-mentioned asymptotically exact mechanism of instanton--anti-instanton proliferation, which is tantamount to the arguments presented in the original papers by Kosterlitz and Thouless.
 A ``strong-disorder" alternative seems therefore unlikely.
Nevertheless, Altman {\it et al.}, inspired by the 1D-specific classical-field mechanism of destroying
global superfluid stiffness by anomalously rare but anomalously weak links,
speculated that an alternative strong-disorder scenario does exist \cite{Altman}.
To corroborate their idea, the authors employed a real-space RG treatment. It is important to realize, however, that the treatment of \cite{Altman} is essentially uncontrolled,
abandoning the usual LL paradigm in favor of the ``Coulomb blockade" single-particle nomenclature promoted to macroscopic scales.

This, in turn, was countered by the theorem of critical self-averaging, which implies that the LL picture holds at criticality \cite{2013}. In combination with the Kane--Fisher result that a single weak link is an irrelevant
perturbation at $K > 1$, this seemed to leave no room for alternatives to the GS scenario because no other asymptotically exact mechanisms for destruction of superfluidity were known. However, recently three of us have realized \cite{2014} that previous studies have overlooked the difference in the outcome of the
Kane-Fisher renormalization for weak links, which occur with finite probability per unit
length, relative to the one for a single link in an infinite system.
The difference is in the classical-field mechanism of suppressing the superfluid stiffness by weak links
(for a discussion, see \cite{RMP,2013}): In  {\it absolute} units, the Kane-Fisher renormalization
is always towards making links weaker---and the weaker the link, the stronger its effect on $\Lambda$.
Despite the fact that at $K>1$, a single weak link cannot destroy superfluidity, the combined
effect of all anomalously weak links in suppressing $\Lambda$ may be strong enough
at large system sizes to drive the system into an insulating state at $K_c > 3/2$, when
the generic instanton--anti-instanton pair dissociation mechanism remains irrelevant.
Note that long-wave hydrodynamic phonons are crucial for this scenario to work---this is in sharp contrast
with the real-space RG treatment of \cite{Altman}.

The new weak-link universality class is termed ``scratched-XY criticality"  because, under the
standard 1D-quantum-to-2D-classical mapping, it corresponds to the superfluid-normal phase transition
in the 2D classical XY model with correlated disorder in the form of parallel ``scratches" (Josephson barriers).
At high enough disorder strength, the scratched-XY (sXY) criticality preempts the BKT transition.
The crucial quantity behind the sXY criticality is the exponent $\zeta$, characterizing the scaling $\propto 1/N^{1-\zeta}$
of the weakest link value found among the $N \gg L$ disorder realizations in a system of fixed mesoscopic size $L$.
The exponent $\zeta$ is of microscopic nature in the sense that it does not involve the Kane--Fisher renormalization of the links/scratches, nor the system size.
[In the 2D classical case, the analog of the Kane--Fisher renormalization is the renormalization due to the thermal fluctuations of the phase field across the scratch.]
The most spectacular manifestation of the special role played by $\zeta$ in sXY criticality is the relation \cite{2014}
\begin{equation}
K_c=1/\zeta .
\label{crit_rel}
\end{equation}

Since the size of the link gets progressively larger for weaker links
(if the barrier height is limited, its length has to increase)
the RG flow is formulated in the space of system sizes rather than  distances.
When doubling the system size, $L \to 2L$, the renormalization of the macroscopic superfluid stiffness is due to one or a few anomalously weak links. Also, links driving the flow at a given system size $L$ are
anomalously rare in the sense that these links are statistically insignificant
({\it i.e.,} absent in the vast majority of realizations) for system sizes much smaller than $L$.

In the present work, we  focus on solving two outstanding problems of the sXY criticality:
(i) a qualitative and quantitative understanding of the interplay between the
sXY and GS scenarios in the vicinity of corresponding tri-critical point
on the SF-BG phase boundary, and
(ii) applying the theory to the disordered Bose-Hubbard model
\begin{equation}
H= - \sum_{\langle i j \rangle} \left(  a_i^\dag a_j^{\,} + h.c. \right) + \frac{U}{2} \sum_{i} {n_i(n_i-1)} + \sum_{i} {(\varepsilon_i\! -\! \mu)n_i}\, .
\label{BH}
\end{equation}
Here,  $a_i^\dag\, (a_i^{\,})$ is the boson creation (annihilation) operator for the site $i$, $\langle \cdots \rangle$ denotes nearest-neighbor sites, $n_i=a_i^\dag a_i$ is the density operator, and $\varepsilon_i$ is the random on-site potential uniformly distributed on the interval $[-\Delta,\Delta]$ with no spatial correlation. Without loss of generality, the hopping amplitude is set equal to unity.  We confine ourselves to  unit filling factor.

Given the exponential divergence of the correlation length on approach to the quantum critical point,
brute-force numerical and experimental approaches to reveal the thermodynamic limit properties are out of the question.
In the vicinity of the critical point, the system size $L$ should be considered as another ``parameter."
We thus look for an asymptotically exact RG theory describing physics at large enough $L$,
thereby  allowing one to analytically extrapolate numerical and experimental data to arbitrarily large $L$.

The mutual effect of sXY and GS criticalities is described by three flow equations. Our analysis of the flow reveals an interesting fact that the competition is unequal: The  GS (instanton--anti-instanton) contributions remain subdominant  with respect to those of the weak links till the tri-critical point; {\it i.e.}, physics of weak links defines
the shape of the SF-BG phase boundary in the vicinity of this point. The developed theory allows us to interpret the results of our extensive quantum Monte Carlo  simulations of the model (\ref{BH}) by fitting the finite-size data with the flow equations. An integral part of the numeric analysis is extracting the exponent $\zeta$ from statistically rare disorder realizations in samples of moderate sizes. We find that the part of the SF-BG phase boundary
controlled by the sXY universality class is rather significant---comparable in size to the part described by the Giamarchi--Schulz scenario.

The paper is organized as follows. In section \ref{sec:II}, we render the theory leading to the set of three flow equations. In section \ref{sec:III}, we review
basic results for the sXY universality class.  In section \ref{sec:IV}, we analyze the interplay between sXY and GS criticalities.
In section \ref{sec:V}, we apply the theory to the model (\ref{BH}) at unit filling.

%%%%%%%%%%%%%%%%%%%%%%%%%%%%%%%%%%%%%%%%%%%%%%%%%%%%%%%%%%%%%%%%%%%%%
\section{Asymptotically exact theory of superfluid-insulator transition}
\label{sec:II}
%%%%%%%%%%%%%%%%%%%%%%%%%%%%%%%%%%%%%%%%%%%%%%%%%%%%%%%%%%%%%%%%%%%%%

%--------------------------------------------------------------%
\subsection{Weakest links}
%--------------------------------------------------------------%
Even in a classical-field 1D system at zero temperature, weak links can drive a phase transition from a superfluid to a peculiar state with zero macroscopic superfluid stiffness~\cite{RMP}.
Such links are anomalously rare but anomalously weak. Theories addressing such links typically assume a strong power-law distribution for a typical weakest link  in a system of length $L$ as
\begin{equation}
J_0^{(L)} \propto { \frac{1}{L^{1-\zeta}} },
\label{J_0_L}
\end{equation}
where $J_0^{(L)}$ can be thought of as a Josephson coupling over the weak link, and $\zeta$ is a well-defined, irrenormalizable microscopic property of the disordered model. Equation (\ref{J_0_L}) can be justified microscopically
in a number of situations~\cite{RMP}, and, in particular, for the classical-field counterpart of the model
(\ref{BH}).

We postulate that (\ref{J_0_L}) is valid for our quantum case.  Specifically, the building blocks of the RG theory are
patches of LL separated by sharply defined weak links, the nature of the latter being
qualitatively similar to weak links in the classical-field system.
By ``sharp" we mean the absence of logarithmic corrections to the power law distribution,
irrespective of weak-link properties ({\it e.g.}, such as its size).
Under these assumptions, the above-mentioned classical-field phase transition~\cite{RMP}
happens at $\zeta=0$. We will verify in section \ref{sec:V} that (\ref{J_0_L}) indeed
holds numerically in all relevant regimes of the Bose-Hubbard model.

While it is difficult to rigorously derive (\ref{J_0_L}) for the model (\ref{BH}), we can still argue
why one should expect this law (the so called exponentially rare--exponentially weak consideration).
Introduce the {\it typical length}, $r^{(L)}$, of the weakest link in the system of size $L$. In the model (\ref{BH}) and similar systems, the weakest link is nothing but a rare disorder realization such that within the length $r^{(L)}$ we have a mesoscopic piece of an insulating state. With respect to its local superfluid environment, this piece behaves as a Josephson junction,
\begin{equation}
J_0^{(L)} \propto e^{ - c_2 r^{(L)} } ,
\label{J_0_l_0}
\end{equation}
with a certain well-defined (in the limit of $r^{(L)}  \to \infty$ and $L \to \infty$) parameter $c_2> 0$.

Note that $r^{(L)}$ in (\ref{J_0_l_0}) is a sharply defined quantity, since its absolute uncertainty, of the order of some microscopic length, significantly changes the value of
$J_0^{(L)} $.  With this fact in mind, we write the condition for the link in question to happen with a probability of order one as
\begin{equation}
L \, {\rm e}^{- c_1 r^{(L)}}  \sim {\rm const} \, .
\label{prob_l}
\end{equation}
The exponential factor, with a certain parameter $c_1>0$, is justified by the natural requirement that the disorder correlation radius is
much smaller than $r^{(L)}$. The factor $L$ takes into account the number of independent realizations (translations) of the rare region.
From (\ref{prob_l}) we have $ r^{(L)} = (1/ c_1)\, \ln L   +{\cal O}(1)$. Substituting this into  (\ref{J_0_l_0}) yields (\ref{J_0_L}), with $\zeta = 1-c_2/c_1$.

%----------------------------------------------------------------------------------%
\subsection{Kane--Fisher renormalization of the weakest links}
%---------------------------------------------------------------------------------%
The Kane--Fisher renormalization of weak links by hydrodynamic phonons can be studied by evoking Popov's hydrodynamic action over the phase field $\Phi(x, \tau)$ when mapping the quantum problem onto a $(1+1)$-dimensional classical system,
\begin{equation} \label{eq:Popov}
S[\Phi] = \int dx d \tau \left[ i n_{0}(x) \partial_{\tau} \Phi + \frac{\Lambda}{2} \left( \partial_{x} \Phi \right)^{2} + \frac{\kappa}{2} \left( \partial_{\tau} \Phi \right)^{2} \right] ,
\end{equation}
where $n_{0}$ is the expectation value of the local density, which has a spatial dependence originating from the external potential and disorder. The first term is of topological nature and the only non-zero contribution comes from instanton--anti-instanton pairs [vortex--anti-vortex pairs in phase field $\Phi(x,\tau)$]. It is through this term that the external potential and disorder get coupled with instanton--anti-instanton pairs, which in turn renormalize the superfluid stiffness \cite{Kashurnikov96}.

The coupling across a link of strength $J_{0}$ is described by the term
\begin{equation}
J_0 \int d \tau \cos[ \Phi_{+}(\tau) - \Phi_{-}(\tau)],
\label{weak_link_term}
\end{equation}
to be added to Popov's hydrodynamic action with $\Phi_{\pm}$ being the phase field just before and after the link. The link is considered weak if the condition
\begin{equation}
J_0 \lambda_0 \kappa \ll  1
\label{weak_cond}
\end{equation}
is met at the microscopic cutoff scale $\lambda_0$.
Under this condition, the term (\ref{weak_link_term}) is perturbative with respect to the rest of the hydrodynamic action consisting of two independent parts separated by the link. The shortest way to derive (\ref{weak_cond})
is to consider two systems of microscopic size $\sim \lambda_0$ connected by the link $J_0$ at $T=0$. The link is perturbative if the number of particles in each of the two systems is a good quantum number despite the presence of the link. Speaking in the ``Coulomb-blockade" language, this requires that $J_0$ be much smaller than
the charging energy. Recalling that the  latter is $\sim 1/ (\kappa \lambda_0)$, we get (\ref{weak_cond}).

For a weak link, Kane and Fisher demonstrated that one can integrate out short wavelength phonons perturbatively.
A remark is in order here concerning a potential danger of the condition $ r^{(L)} \gg \lambda_0$ taking place for
rare weak links. In the perturbative renormalization, the fields $\Phi_{+}$ and $\Phi_{-}$ in
(\ref{weak_link_term}) are considered as independent and the spatial distance between them has no
effect on the final result.

The Kane-Fisher flow equation for the renormalized strength of a weak link reads (we do not rescale distance for later convenience)
\begin{equation} \label{eq:KF}
\frac{d J(\lambda)}{d \ln \lambda} = - \frac{1}{K(\lambda)} J(\lambda) \, .
\end{equation}
The critical condition $K_c^{\rm KF}=1$ for the physics of a single weak link with fixed microscopic strength in an otherwise homogenous LL follows then directly from the scaling dimension of $J$: For $K< 1$ the link cuts the system into halves [$J (\lambda ) \lambda \kappa \to 0$, cf.~(\ref{weak_cond})], whereas  the link ultimately becomes strong
$[J(\lambda) \lambda \kappa \sim 1]$ for $K > 1$. Since in a generic disordered system the microscopic strength of the typical weakest link is also changing under the RG flow, it is instructive to consider how the Kane-Fisher renormalization works in the superfluid phase at intermediate length scales rather than just taking the thermodynamic limit. For the sXY-criticality to occur, the critical LL parameter must be $K_{c} \ge 3/2$, meaning that while the absolute strength of the link, $J(\lambda)$, decreases under the RG flow, its relative strength $J\lambda \kappa$ increases. Hence, the renormalization of $J$ inevitably stops at the \emph{clutch scale} $\lambda_{*}$ given by
\begin{equation}
J(\lambda_{*}) \lambda_{*} \kappa  \sim 1 \, ,
\label{clutch}
\end{equation}
where perturbation theory is no longer valid.  Therefore, upon completing the Kane-Fisher renormalization at the clutch scale $\lambda_{*}$, the link (\ref{J_0_L}) picks up a   certain renormalization factor $f(\lambda_{*})$. With the help of the integral form of (\ref{eq:KF}) this factor can be expressed as
\begin{equation}
f(\lambda_{*}) = \exp \left[ -\int_0^{\ln \lambda_{*}} \frac{d \ell}{K(\ell)}  \right] , \qquad \ell = \ln (\lambda / \lambda_0) \, .
\label{KF_factor}
\end{equation}

The clutch condition (\ref{clutch})  yields the following relation between $\lambda_*$ and the microscopic strength of the weak link (below $J_0\equiv J_0^{(L)}$)
\begin{equation}
J_0 f(\lambda_{*})  \lambda_{*} \kappa  \sim 1 \, .
\label{clutch2}
\end{equation}
Finally, the renormalization of the superfluid stiffness by the weakest link,
$J_* = J_0 f(\lambda_{*})$, in the system of size $\sim L$ obeys the flow equation
\begin{equation}
\frac{d \Lambda^{-1}}{d \ell} \sim  \frac{1}{J_*  L}  \sim \frac{1}{J_0 f(\lambda_{*}) \, L}  \sim \kappa \frac{\lambda_*}{L}\, .
\label{stiffness_flow}
\end{equation}
To avoid potential problems with the tail of the distribution of (abnormally) weak links, we understand $\Lambda^{-1}(\ell)$ as the median
value for different disorder realizations at a given system size. The theorem of critical self-averaging \cite{2013} allows us to deal with the median value rather than the whole distribution.

Since $\kappa$ does not flow with $\ell$, we readily rewrite (\ref{stiffness_flow}) as the flow equation for $K$
\begin{equation} \label{eq:RG_Kw}
\frac{d K}{d \ell} = - w K^{3} \, ,
\end{equation}
where
\begin{equation}
w \sim \,\lambda_* / L \equiv e^{\ell_{*} -\ell} \,,
\label{w_definition}
\end{equation}
and $\ell_{*} = \ln( \lambda_*/ \lambda_0)$, $\ell = \ln (L/\lambda_0)$. Along with $K(\ell)$, the function $w(\ell)$ (characterizing the role of weak links at a given $\ell$) plays a central role
in the flow equations. The controllability of the RG theory requires that $d K(\ell) /d \ell  \ll K(\ell) $ [{\it i.e.,} $K(\ell)$ flows slowly along the scales of distance], which translates into
\begin{equation}
w(\ell) K^{2}(\ell) \ll 1\, .
\label{w_inequality}
\end{equation}
Below we will see that this requirement is consistent with the flow equations since the latter guarantee $\lim_{\ell \to \infty} w(\ell) =0$ in the superfluid phase, including the critical point.

To obtain the flow equation for $w(\ell)$, we substitute $J_0$ of (\ref{J_0_L}) into (\ref{clutch2}),
\begin{equation}
\frac{\lambda_{*}  }{L^{1-\zeta}} \,  f(\lambda_{*}) \, = \, {\rm const} \, .
\label{eq:clutch0}
\end{equation}
Taking the logarithm on both sides and differentiating with respect to $\ell$ using (\ref{KF_factor}),
\begin{equation} 
\frac{d \ell_{*}(\ell )}{d \ell}  \, = \,   \frac{1 -\zeta}{1- K^{-1}(\ell_{*})}   \, .
\label{eq:clutch1}
\end{equation}
Differentiating the definition (\ref{w_definition}) with respect to $\ell$ and using (\ref{eq:clutch1}), we get
\begin{equation}
\frac{d w}{d \ell}  \, =  \, \frac{ 1 - \zeta K(\ell_{*}) }{K(\ell_{*}) - 1} \, w \, .
\label{w_flow_1}
\end{equation}

Up to higher-order corrections, we can replace $K(\ell_{*})$ with $K(\ell)$ in the r.h.s. of (\ref{w_flow_1}).
Indeed, Taylor expanding $K(\ell_{*})$ with (\ref{eq:RG_Kw}) and (\ref{w_definition}) taken into account, we have
\begin{equation}
    \begin{split}
        K(\ell_{*}) & = K(\ell) + w K^{3}(\ell) (\ell - \ell_{*}) + \cdots \\
                    & = K(\ell) + K^{3}(\ell) w \ln w^{-1} + \cdots \, ,
    \end{split}
\label{Taylor_K}
\end{equation}
so that,when writing $K(\ell) = \zeta^{-1} + x(\ell)$,  the r.h.s. of (\ref{w_flow_1}) becomes 
\begin{equation}
\frac{1 -\zeta K(\ell_{*})}{K(\ell_{*}) -1 } \, w \, = \, -\frac{\zeta [x(\ell) + K^{3}(\ell) w \ln w^{-1} + \cdots ] }{K(\ell) + K^{3}(\ell) w \ln w^{-1} -1 + \cdots} \, w \, .
\end{equation}
Replacing $K(\ell_{*})$ with $K(\ell)$ is legitimate if $K^{3}(\ell) w \ln w^{-1} \ll x(\ell)$.
For large enough $\ell$, this condition is always satisfied.
As we will see later [from (\ref{eq:cri_xw1}) and (\ref{eq:cri_xw2})], even in the worst case scenario, {\it i.e.} at the critical point when the asymptotic flow of $K$ is the strongest and $x(\ell) \to 0$,
we have
 \begin{equation}
K^3 w \ln w^{-1}  \, \propto  \,  \frac{1}{\zeta^{2}\ell^{2}} \ln \frac{\ell}{\sqrt{\zeta}} \, \ll \, \frac{1}{\zeta^{2} \ell} \, \sim \, x(\ell).
\end{equation}

Finally, we consider the renormalization of $K$ by instanton--anti-instanton pairs
(for details see, {\it e.g.}, \cite{Kashurnikov96}). This brings us to
three coupled RG equations
\begin{align}
    {\frac{dy}{d \ell} } & =  { (3/2-K)\, y } \, , \label{eq:RG_a} \\
    {\frac{dK}{d \ell} } & =  { -K^2 y^2 -K^3 w } \, ,\label{eq:RG_b} \\
    {\frac{dw}{d \ell} } & =  {\frac{1-\zeta K}{K-1} \, w} \, . \label{eq:RG_c}
\end{align}
Equation (\ref{eq:RG_a}) is the standard Kosterlitz--Thouless equation for the flow of the vortex fugacity $y$, except that the coefficient in front of $y$ is $(3/2 - K)$ instead of $(2-K)$. This is because in the $(1+1)$D representation of the 1D disordered system only vertical vortex--anti-vortex pairs contribute to the renormalization of the superfluid density \cite{Kashurnikov96}. Equation (\ref{eq:RG_b}) describes the renormalization of the LL parameter by the instanton--anti-instanton term $y^{2}$ and by the weak-link term $w$. Finally, equation
(\ref{eq:RG_c}) is the same as (\ref{w_flow_1}), with the above-discussed replacement $K(\ell_*) \to K(\ell)$.

Once the RG flow hits the point $K(L) = \max\left\{ 3/2, \zeta^{-1}\right\}$, it quickly flows to an insulating state, $K(\infty)=0$. We thus identify two different universality classes: (i) the GS universality class  with the universal critical value of the LL parameter $K_{c} = 3/2$, and (ii) the sXY universality class, where the critical LL parameter is semi-universal, $K_{c}=\zeta^{-1}$.

%%%%%%%%%%%%%%%%%%%%%%%%%%%%%%%%%%%%%%%%%%%%%%%%%%%%%%%%%%%%%%%%%%%%%
\section{Physics in the weak-link regime}
\label{sec:III}
In this section, we study the critical behavior in the weak-link regime where the phase transition to the insulating phase is driven by the weakest links whereas proliferation of instanton--anti-instanton paris remains irrelevant. Explicit solutions to the flow equations are obtained to demonstrate the BKT-like nature of the criticality.  Equation (\ref{stiffness_flow}) takes into account only the contributions of
{\it microscopic} weakest  links. We justify this crucial assumption by showing that the contribution of {\it composite} weak links is always subdominant.

%%%%%%%%%%%%%%%%%%%%%%%%%%%%%%%%%%%%%%%%%%%%%%%%%%%%%%%%%%%%%%%%%%%%%
%%%%%%%%%%%%%%%%%%%%%%%%%%%%%%%%%%%%%%%%%%%%%%%%%%%%%%%%%%%%%%%%%%%%%
%--------------------------------------------------------------%
\subsection{Criticality driven by the weakest links}
%--------------------------------------------------------------%
%
We define the weak-link regime by the requirement $\zeta < 2/3$ so that the criticality is due to the weakest links. To study this critical behavior we neglect the instanton--anti-instanton pair term $y^2$ in (\ref{eq:RG_b}) and the relevant RG equations simplify to
\begin{align}
    \frac{dK}{d \ell}  &= {-K^3 w } \, ,\label{eq:RG_wk_1} \\
    \frac{dw}{d \ell}  &= {\frac{1/\zeta-K}{(K-1)/\zeta} \, w} \, .\label{eq:RG_wk_2}
\end{align}
The first integral is  found by dividing (\ref{eq:RG_wk_2}) through (\ref{eq:RG_wk_1}),
\begin{equation}
dw = \frac{\zeta K-1}{(K-1)K^{3}} d K \, ,
\end{equation}
and then integrating both sides of the equation. It reads
\begin{equation} \label{eq:wk_fst_int}
w = A - f(\zeta, K) \, ,
\end{equation}
where $A$ is an integration constant depending on the microscopic parameters and
\begin{equation} \label{eq:f_func}
f(\zeta,K) = \frac{1}{2 K^{2}} + \frac{1-\zeta}{K} + (1-\zeta) \ln \frac{K-1}{K}.
\end{equation}
Since on approach to the critical point from the superfluid phase $w(\ell = \infty)=0$,  the integration constant $A$ satisfies
\begin{equation} \label{eq:A_f}
A = f(\zeta, K(\ell = \infty))
\end{equation}
in the superfluid phase and at the critical point. In our model (\ref{BH}), $A = A(U, \Delta)$. Let us follow the parameter $A$ along a line segment in the plane $(U, \Delta)$. Parameterizing the segment as
$U=U(t)$, $\Delta=\Delta(t)$, with  a certain parameter $t$, and assuming that the segment crosses the SF-BG critical line at the point $(U_{c}, \Delta_{c})$, in the vicinity of the point 
$(U_{c}, \Delta_{c})\equiv (U(t_c), \Delta(t_c))$ we have:
\begin{equation}
A(t) \approx A(t_c) + A_{1} (t_{c}- t) \, .
\label{A_of_t}
\end{equation}
Similarly, following $\zeta$ on the same line segment we have
\begin{equation}
\zeta (t) \approx \zeta (t_c) + A_{2} (t_{c}- t)\, .
\label{zeta_of_t}
\end{equation}
Here $A_1$ and $A_2$ are certain constants, $t_{c}$ is the critical value of $t$. Without loss of generality, we assume $t<t_{c}$ in the superfluid phase.   
Substituting (\ref{A_of_t}) and (\ref{zeta_of_t}) into (\ref{eq:A_f}) and keeping only the leading terms, we get
\begin{equation}
K_{\infty}(t) -  \zeta^{-1} (t_c) \,  \propto \, \sqrt{t_{c}-t} \, .
\end{equation}
Hence, the LL parameter in the scratched-XY universality class demonstrates the same square-root cusp as in the conventional BKT case. The analogy with BKT transition can be traced further 
by investigating the linearized flow equations near the critical point:
\begin{align}
    \frac{d \tilde{w}}{d \ell} &= -2 \tilde{x} \tilde{w} \, ,\label{eq:tilde_xw_1}\\
    \frac{d \tilde{x}}{d \ell} &= - \tilde{w} \, .\label{eq:tilde_xw_2}
\end{align}
Here we first introduce $x(\ell)$ such that $K(\ell)=\zeta_{c}^{-1} + x(\ell)$, and then rescale $x$ and $w$:
\begin{align}
    \tilde{x}(\ell) & = \dfrac{\zeta_{c}^{2} \; x(\ell)}{2(1-\zeta_{c})} \, , \\
    \tilde{w}(\ell) & = \dfrac{w(\ell)}{2(1-\zeta_{c})\zeta_{c}} \, .
\end{align}
The first integral of the system (\ref{eq:tilde_xw_1})--(\ref{eq:tilde_xw_2}) is readily found:
\begin{equation} \label{eq:tilde_xw_int}
\tilde{w}(\ell) = \tilde{x}^{2}(\ell) - \tilde{A} \, .
\end{equation}
Here $\tilde{A}$ is an integration constant. By definition, $\tilde{A}$ vanishes at the critical point.   The constant $\tilde{A}$ is an analytic function of the microscopic parameters, because (\ref{eq:tilde_xw_int}) is valid for {\it finite} $\ell$ where both $\tilde{x}(\ell)$ and $\tilde{w}(\ell)$ cannot have singularities. This enables us to expand $\tilde{A}$ as  
\begin{equation}
\tilde{A} \approx B (t_{c} - t) \, ,
\end{equation}
with a positive constant $B$. 

The form of the complete solutions depends on the sign of $\tilde{A}$. For the superfluid side ($\tilde{A}>0$), we have:
\begin{align} 
    x(\ell)  & = \frac{2(1-\zeta_c)}{\zeta_c^{2}} \frac{\sqrt{\tilde{A}}}{\tanh \left[ \sqrt{\tilde{A}}\; (\ell - \ell_{0}) \right]}\, , \label{eq:sf_xw1} \\
    w(\ell)  & = \frac{2(1-\zeta_c)\zeta_c \tilde{A}}{\sinh^{2} \left[ \sqrt{\tilde{A}}\; (\ell - \ell_{0}) \right]} . \label{eq:sf_xw2}
\end{align}
For the insulator side ($\tilde{A}<0$), the solution is:
\begin{align} 
    x(\ell)  &= \frac{2(1-\zeta_c)}{\zeta_c^{2}} \frac{\sqrt{|\tilde{A}|}}{\tan \left[ \sqrt{|\tilde{A}|}\; (\ell - \ell_{0}) \right]}\, , \label{eq:insu_xw1} \\
    w(\ell)  &=  \frac{2(1-\zeta_c)\zeta_c |\tilde{A}|}{\sin^{2} \left[ \sqrt{|\tilde{A}|}\; (\ell - \ell_{0}) \right]} . \label{eq:insu_xw2}
\end{align}
At the critical point, we have
\begin{align} 
    x(\ell)  &= \frac{2(1-\zeta_c)}{\zeta_c^{2}} \frac{1}{\ell - \ell_{0}}\, , \label{eq:cri_xw1} \\
    w(\ell)  &= \frac{2(1-\zeta_c)\zeta_c}{\left(\ell - \ell_{0} \right)^{2}} \, . \label{eq:cri_xw2} 
\end{align}
The second integration constant $\ell_{0}$ has a trivial meaning of the logarithm of the length unit.
 The correlation length $\xi$ hence diverges near the critical point in the same characteristic way as in the conventional BKT case:  $\ln \xi  \sim 1/\sqrt{|\tilde{A} |} \sim 1/\sqrt{|t_c - t|}$.

%--------------------------------------------------------------%
\subsection{Irrelevance of composite weak links}
%--------------------------------------------------------------%

To make sure that the theory of the sXY universality is fully consistent, it is important to demonstrate that {\it composite} weak links play only a subdominant role
(in contrast to the assumptions of \cite{Altman}) and thus can safely be ignored in the RG analysis.
Consider the simplest composite weak link---to be referred  as a $d$-pair for brevity---formed
by two microscopic  weak links (of comparable microscopic strength $J_0$) separated by a distance $d$
much larger than the microscopic scale but much
smaller than the clutch scale for any of the two links. Up to the length scale $\lambda \sim d$, the Kane--Fisher renormalization of the two links takes place
independently and reduces to multiplying  each of the two $J_0$'s by the factor
\begin{equation}
f(d ) = \exp \left[-\int_0^{\ln d} { \frac{d \ell}{K(\ell)} } \right]  \, ,
\label{RGfactor}
\end{equation}
Mathematically, the merger of two renormalized links of strength $J_1(d) \sim J_2(d) \sim J_0 f (d)$ into one composite link
takes place upon integrating out the phase field between the two links. [We note in passing that, in the renormalized theory, the phase field between the links depends only on $\tau$ and not on $x$.] An explicit calculation yields the following estimate for the effective strength of the composite link
\begin{equation}
J_{\rm comp}^{(d)} \sim J_1(d) J_2(d) \kappa d .
\label{J_comp}
\end{equation}
This estimate is physically transparent, and immediately follows from second-order
perturbation theory by considering the $d$-pair as a quantum dot in the ``Coulomb blockade" regime.
Then, $J_1(d)$ and  $J_2(d)$ are two effective single-particle tunneling matrix elements and $1/(\kappa d)$ is the charging energy of the dot.
For length scales $\lambda > d$ the Kane--Fisher renormalization of the composite link
reduces to multiplying (\ref{J_comp}) by the factor
\begin{equation}
\exp \left[-\int_{\ln  d}^{\ln \lambda} { \frac{d \ell}{K(\ell)} } \right] \, =\, f(\lambda) / f(d) \, .
\label{RGfactor2}
\end{equation}
Hence, for the renormalized strength of a $d$-pair we have
\begin{equation}
J_{\rm comp}^{(d)} (\lambda) = \kappa d J_0^2 f(d) f(\lambda) \qquad \quad (\lambda > d)\, . \quad
\label{d_pair}
\end{equation}
The clutch scale for the $d$-pair, $\lambda_*^{(d)}$, then follows from the condition $J_{\rm comp}^{(d)} (\lambda) \lambda_*^{(d)} \sim 1 $
(in units $\kappa$ =1)
\begin{equation}
J_0^2 f(d) f(\lambda_*^{(d)} ) \, \lambda_*^{(d)} d\sim 1 \, .
\label{clutch_pair}
\end{equation}
This is all we need to compare the contribution of $d$-pairs  with the one of isolated weak links.

For a given system size $L$ and scale $d$  we only need to account for those $d$-pairs which occur with a probability of order one because pairs with higher density are absorbed into the renormalized value of
$\Lambda_s (L)$ whereas unlikely pairs are accounted for at larger system sizes.
This defines the characteristic $J_0$ as a function of $L$ and scale $d$:
\begin{equation}
\left[ J_0^{1/(1-\zeta)}\right]^2 d = 1/L \, 
\label{probability1}
\end{equation}
for $d$-pairs within the scale $d$. For an explicit calculation, we confine ourselves to the critical point, which is the most dangerous regime for the putative relevance
of composite weak links. Here, with (\ref{eq:cri_xw1}) we have
\begin{equation}
f(\lambda ) =  \lambda^{-\zeta} \ln^{2(1-\zeta)}(\lambda) \qquad \quad \mbox{(critical flow)} \, .
\label{RGfactor_crit}
\end{equation}
Combining (\ref{clutch_pair}) and (\ref{probability1}) with (\ref{RGfactor_crit}), we obtain
\begin{equation}
\left[ \ln^2(d) \ln^2(\lambda_*^{(d)}) \; \frac{\lambda_*^{(d)}}{L} \right]^{1-\zeta} \sim 1 \, .
\label{pairdone}
\end{equation}
Finally, replacing $\lambda_*^{(d)}$ with $L$ under the logarithm,
we get (with logarithmic accuracy)
\begin{equation}
\frac{\lambda_*^{(d)}}{\lambda} \sim \frac{1}{ \ln^2\! d \, \ln^2\! L} \;,
\label{pairdone2}
\end{equation}
and observe that the contribution of large $d$-pairs to the renormalization of $\Lambda$
is suppressed relative to (\ref{eq:cri_xw2}) by a factor of $\ln^{-2}(d)$.
Most importantly, the integral over the pair scales
$\int d[\ln d] $---yielding the total renormalization contribution of all relevant $d$-pairs---converges
at the {\it lower} limit, where microscopic pairs (and other multi-link complexes)
are an integral part of the original exponentially-rare exponentially-weak distribution
of single links.

%%%%%%%%%%%%%%%%%%%%%%%%%%%%%%%%%%%%%%%%%%%%%%%%%%%%%%%%%%%%%%%%%%%%%
\section{Interplay between weak-link and Giamarchi--Schulz scenarios}
\label{sec:IV}
%%%%%%%%%%%%%%%%%%%%%%%%%%%%%%%%%%%%%%%%%%%%%%%%%%%%%%%%%%%%%%%%%%%%%

The next natural question to ask is how the well-known GS criticality based on proliferation
of instanton--anti-instanton pairs crosses over to the weak-link criticality.
It turns out that weak links are more aggressive and the instanton--anti-instanton pairs can be
neglected in the asymptotic flow at the tri-critical point. A direct consequence of this fact
is the continuous first-order derivative of the transition line at the tri-critical point.

%--------------------------------------------------------------%
\subsection{RG equations for small deviations from the tri-critical point}
%--------------------------------------------------------------%
To study the competition between weak links and instanton--anti-instanton pairs in the vicinity of the tri-critical point, we rewrite identically $K(\ell)=3/2+x(\ell)$, $ \zeta=2/3+\delta$, and consider $x>0$ and $\delta$ as small parameters to simplify the RG equations. The tri-critical point is fixed by $\delta = 0$ with positive/negative $\delta$ corresponding to the GS/sXY criticality, respectively. Expanding (\ref{eq:RG_a}), (\ref{eq:RG_b}) and (\ref{eq:RG_c}) to leading order in $x$ and $\delta$ results in
\begin{align}
    {\frac{dy}{d \ell} } &= { -xy } \, , \label{eq:RG_linear_a} \\
    {\frac{dx}{d \ell} } &= { -y^2 -w} \, , \label{eq:RG_linear_b} \\
    {\frac{dw}{d \ell} } &= {-\left(\frac{4}{3} x+3 \delta \right) w} \, . \label{eq:RG_linear_c}
\end{align} 
Here functions $y$ and $w$ were rescaled to eliminate multiplicative constants. This is a set of three coupled, first-order differential equations,  implying that the solution will have three free constants. One of them, $\ell_0$, has the same meaning as 
in Sec.~\ref{sec:III} [see (\ref{eq:sf_xw1})--(\ref{eq:cri_xw2})]. Assuming corresponding choice of length units,  we set $\ell_{0}=0$ from now on. 
The other two integration constants, denoted as $C$ and $D$ below, are related to the microscopic
parameters of the system and dictate the location of the SF-BG transition.

Extracting $x$ from (\ref{eq:RG_linear_a}) and plugging it into (\ref{eq:RG_linear_c}) leaves us with
\begin{equation}
\frac{d \ln w}{d \ell} + 3 \delta = \frac{4}{3} \frac{d \ln y}{d \ell} \, .
\end{equation}
Integrating both sides over $\ell$ leads to the first integral
\begin{equation} \label{eq:yw}
y^2=\frac{1}{C^{3/2}} w^{3/2} \, e^{9\delta \ell/2} \, .
\end{equation}
In the weak-link regime, and at the tri-critical point, where $\delta \le 0$, (\ref{eq:yw})
implies $y^{2} \ll w$; \emph{i.e.}, starting from some mesoscopic length scale the weak-link term dominates over the instanton--anti-instanton pairs. The $3\delta$ in (\ref{eq:RG_linear_c}) results in a non-universal critical parameter $K_{c} > 3/2$ since otherwise $w$ diverges. In the GS regime, the flow starts with $y^{2} \ll w$ but ultimately
crosses over to $y^{2} \gg w$ at some length scale $\tilde{\ell}$ (see below) leading to a universal critical LL
parameter $K_{c}=3/2$ and the familiar BKT critical behavior.

Dividing (\ref{eq:RG_linear_b}) by (\ref{eq:RG_linear_c}) and making use of (\ref{eq:yw}), we have
\begin{equation}
\left( 4x/3 + 3\delta \right) dx = \left( C^{-3/2} w^{1/2} e^{ 9 \delta \ell/2} + 1 \right) d w \, .
\end{equation}
Integrating both sides from $\ell_0$ to $\ell$ and utilizing the \emph{second mean value theorem for definite integrals} results in
\begin{equation} \label{eq:xw_eta}
\left( x+ \frac{9	}{4} \delta \right)^2 = \left(\frac{w}{C} \right)^{3/2} e^{9\delta \eta \ell/2} + \frac{3 w}{2}  + D \, ,
\end{equation}
where $0<\eta<1$ and $D$ is our last integration constant.

The meaning of the integration constant $D$ is easily revealed by examining the weak-link critical line where $K_{c}(\infty)=\zeta^{-1}=3/2-9 \, \delta/4$  and $x_{c}(\infty)=-9\,\delta/4$ up to leading order in $\delta$. From (\ref{eq:xw_eta}) the weak-link critical line corresponds to
\begin{equation} \label{eq:wk_c}
D_c=0.
\end{equation}
Since $\delta$ is a microscopic parameter, one can use $\delta$ and $D$ (controlling the location of the tri-critical point and the transition line, respectively) to conveniently parameterize the whole problem.

%--------------------------------------------------------------%
\subsection{Giamarchi--Schulz regime and the parabolic crossover}
%--------------------------------------------------------------%
In the GS regime the critical condition $K_{c}=2/3$ translates into $x_c(\infty) = 0$.
% up to leading order in $\delta$.
Moreover, the GS critical line will lie on the positive side of the $D=0$ line extended beyond the
tri-critical point (see also below). However, as can be seen from (\ref{eq:yw}), the crossover length scale
for the dominance of the vortex fugacity term satisfies the condition $\tilde{\ell} > 1/\delta$, meaning that
at small $\delta$ (i) the initial flow of $K$ is due to weak links, and (ii) the critical line
is closely following the $D=0$ line. Indeed, substituting (\ref{eq:yw}) into (\ref{eq:xw_eta}) gives
\begin{equation} \label{eq:xyw}
\left( x + \frac{9}{4} \delta \right)^{2} = y^{2} e^{-(1-\eta) \delta \ell}  + \frac{3w}{2} + D.
\end{equation}
The condition for the Giamarchi--Schulz critical line is then just
\begin{equation} \label{eq:GS_c}
D_{c} = \left(9 \delta/4 \right)^{2} = \frac{81}{16} \delta^{2} \, ,
\end{equation}
since the first and second terms on the r.h.s flow to zero.

For the 1D disordered Bose-Hubbard model the microscopic parameter $\zeta$ (and thus $\delta$)
and the integration constant $D$ are supposed to be analytic functions of the model parameters $U$ and $\Delta$.
Since (\ref{eq:GS_c}) predicts that the SF-BG boundary has a continuous first derivative across
the tri-critical point in the $(D,\zeta )$-plane, the same property holds in the $(U,\Delta )$-plane.
By the same token the second derivative is discontinuous. Therefore, the crossover is \emph{parabolic}
in the $(U,\Delta )$-plane.

%%%%%%%%%%%%%%%%%%%%%%%%%%%%%%%%%%%%%%%%%%%%%%%%%%%%%%%%%%%%%%%%%%%%%
\section{Ground-state phase diagram of the one-dimensional disordered Bose-Hubbard model}
\label{sec:V}
%%%%%%%%%%%%%%%%%%%%%%%%%%%%%%%%%%%%%%%%%%%%%%%%%%%%%%%%%%%%%%%%%%%%%
\subsection{Protocol of extracting $\zeta$}
Our numerical procedure of extracting $\zeta$ is based on measuring small probabilities, $P \propto J_0^{1/(1-\zeta)}$,
of {\it rare realizations} of disorder, when in a system of (moderate)
size $L$ there is an anomalously weak link, $J_0$, with the clutch scale much larger than the system size:
$\kappa J(L)L \ll 1$. [Here $J(L) \equiv J(\lambda = L)$, see section~\ref{sec:II}.]
With open boundary conditions such a link would cut the system into two essentially independent pieces.
With twisted boundary conditions it acts as a Josephson junction in a superfluid ring:
the particle flux (persistent current) in the ring, $j$, in response to a phase twist, $\varphi$,
is related to $J(L)$ by
\begin{equation}
j = \frac{\partial F}{\partial \varphi} =  J(L) \sin \varphi  \qquad \quad  (T \ll 2\pi^2 \Lambda /L)\, ,
\label{Josephson}
\end{equation}
where $F$ is the free energy. %(and $c$ the sound velocity).
The first equality in (\ref{Josephson}) is absolutely general and does not imply any extra condition.
A delicate aspect of the static thermodynamic response to the gauge phase in low-dimensional systems 
is the necessity to address the contribution of supercurrent  states ({\it i.e.}, states with non-zero
global winding numbers of the phase field around the ring) \cite{PS2000}.
Supercurrent states can dramatically affect the second equality in (\ref{Josephson}) at elevated temperature:
In order to guarantee that their contribution is negligible, we need to consider temperatures much lower
than the energy of the first supercurrent state, $T \ll 2\pi^2 \Lambda /L$.

With (\ref{Josephson}) one relates $J(L)$ to the second derivative of $F$
with respect to $\varphi$ at $\varphi=0$:
\begin{equation}
J(L) = \left. \frac{\partial^2 F}{\partial \varphi^2} \right|_{\varphi=0} \, .
\label{second_der}
\end{equation}
Within the worldline representation (used in our numerical simulations by the worm algorithm),
the r.h.s. of (\ref{second_der}) is readily obtained by the well-known Pollock--Ceperley formula \cite{Pollock}
\begin{equation}
\left. \frac{\partial^2 F}{\partial \varphi^2} \right|_{\varphi=0} = T \left.
\langle M^2 \rangle  \right|_{\varphi=0} \, ,
\label{Pollock-Ceperley}
\end{equation}
expressing the linear response as the variance of the worldline winding number $M$ at a given temperature $T$ at $\varphi=0$.

The measured strength $J(L)$ of the anomalously weak link is, of course,
different from the microscopic value $J_0$ due to Kane--Fisher renormalization $(\lambda_0/L)^{1/K}$.
However,
this renormalization does not depend on $J_0$, and cannot affect the power-law exponent $\zeta$. Hence,
\begin{equation}
J_0  \propto  T \left.  \langle M^2 \rangle  \right|_{\varphi=0} 
 \, 
\label{abnormal_J_0}
\end{equation}
for all weak links at given $L$ and $T$. Furthermore, for purposes of extracting $\zeta$, (\ref{abnormal_J_0}) can be used
even at an elevated temperature $T > 2\pi \Lambda /L$ when the persistent current response
(\ref{Josephson}) is dramatically renormalized (suppressed) by the supercurrent states \cite{PS2000}.
Indeed, in view of the perturbative nature of the weak link response,
both Kane--Fisher and supercurrent  renormalizations reduce to a certain factor
$\tilde{f}(L,T)$ independent of $J_0$. In terms of $J_0$ and $\tilde{f}(L,T)$, we then have
\begin{equation}
J_0 \, \tilde{f}(L,T) = T \left.  \langle M^2 \rangle  \right|_{\varphi=0}  \, ,
\label{abnormal_J_0_exact}
\end{equation}
justifying (\ref{abnormal_J_0}).
%As mentioned previously, any proportionality coefficient in (\ref{abnormal_J_0}) would be irrelevant for determining the power law exponent $\zeta$ (keeping $L$ and $T$ constant).

If the concept of irrenormalizable weak links is correct, then finding the weakest link among the
$N\gg 1$ different disorder realizations in a system of a moderate size $L$ is equivalent to doing
the same in a much larger single system of size $L_1=LN$. This leads to an efficient protocol for determining $\zeta$ using the approach by varying $N$ (and $L$, to validate the concept).
The three natural limitations on $L$ and $N$ are:
(i)  $L$ has to be sufficiently large to capture all the essential microscopic physics of weak links;
(ii) $N$ should not be ``astronomically" large to ensure that the size of the typical weakest link remains much smaller than $L$.
In practice, this condition is hard to violate even for a moderate value of $L=20$;
(iii) $N$ should be large enough to ensure that the clutch scale of the weakest link in a system of size $L$
exceeds $L$.

\begin{figure}[htbp]
\centering
\includegraphics[width=\linewidth]{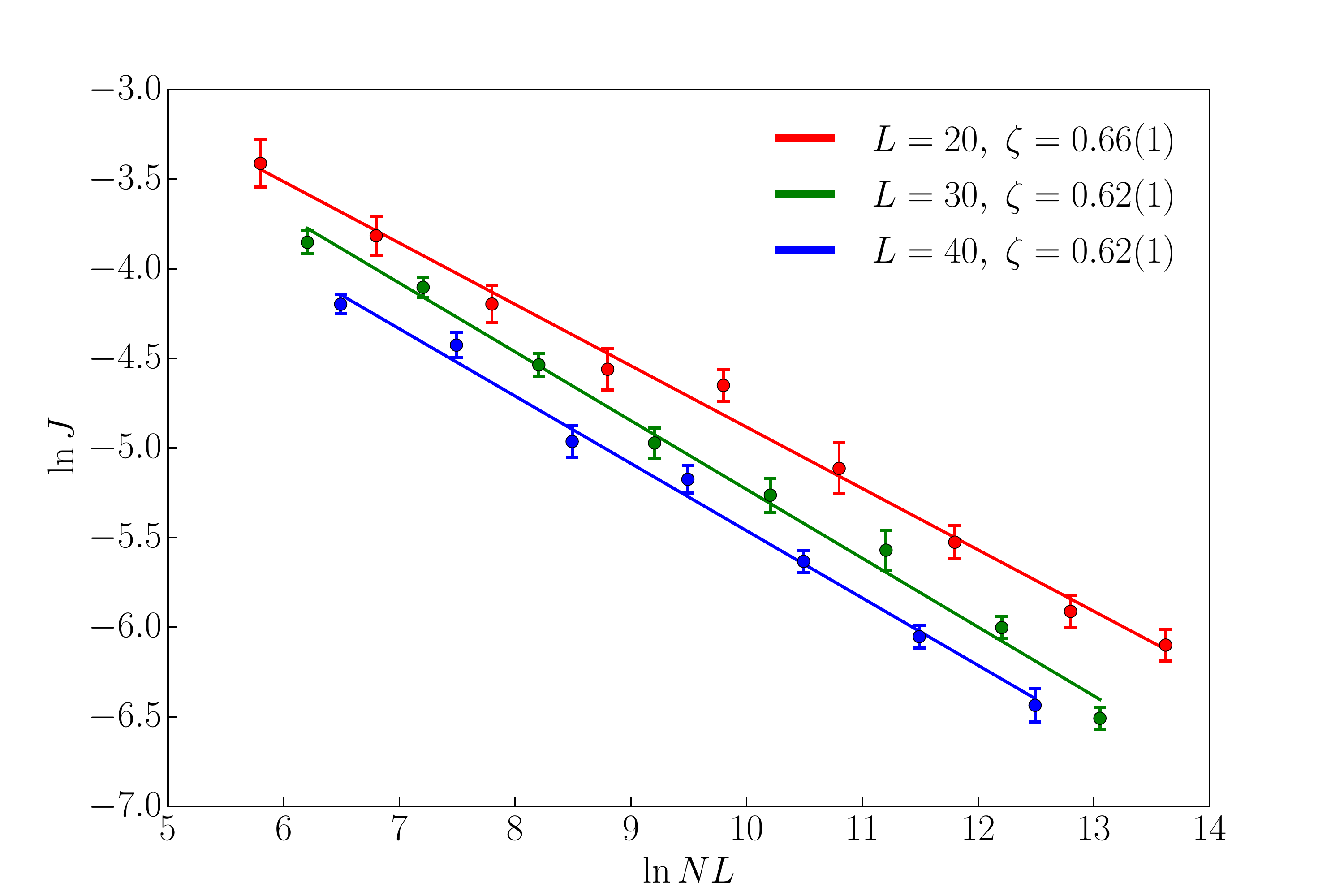}
\caption{Determining $\zeta$ for $(U=4.2, \Delta=3.8)$ using (\ref{loglogfit}).
	The error bars for $\zeta$ denote one standard deviation (deduced by  the confidence level for a linear fit to all data points). }
\label{fig:figure1}
\end{figure}

To extract $\zeta$, we simulate $N_iL=[10e^{i}]$ ($i=1,2,3, \dots$)  different disorder realizations
(here $[...]$ stands for the closest integer) and record the smallest weak link parameter $J_i$
in a given simulation run; {\it i.e.}, we use an equidistant mesh for $\ln NL$ to probe different length scales.
To suppress statistical noise, the procedure is repeated $R=20, 30, 40$ times
(depending on system size) and results from multiple runs, $J_i^{(k)}$ ($k=1,2, \dots ,R$),
are used to determine the typical weak link value as an average over all runs,
$J(N_iL) = \langle J_i^{(k)} \rangle_R$. Its error bar follows from the data dispersion,
$\delta J(N_iL) = \sqrt{ \{  \langle [J_i^{(k)}]^2\rangle_R -J^2(N_iL)\} /R} $.
Finally, the data for $\ln J(NL)$ is fitted to a linear dependence
\begin{equation}
\ln J = (\zeta-1) \ln NL +{\rm const} \;,
\label{loglogfit}
\end{equation}
to extract the power-law exponent $\zeta$.
A characteristic example of the $\zeta$-analysis is shown in figure~\ref{fig:figure1}.
We see that the data is perfectly described by a linear dependence, leading to an accurate
determination of $\zeta$. Within three-$\sigma$ error margins, the
slope of the linear fit does not change when going from $L=20$ over $L=30$ to $L=40$.
This behavior is in perfect agreement with the notion of $\zeta$ as an irrenormalizable
microscopic parameter in the superfluid phase and in the critical region.
In fact, figure~\ref{fig:figure1} is representative of the worst-case
scenario because according to our RG analysis (see figure~\ref{fig:figure3} below)
the parameter set $(U=4.2, \Delta=3.8)$ belongs to the BG phase in close vicinity
of the critical point when $L=40$ is still smaller than the correlation length.

%%%%%%%%%%%%%%%%%%%%%%%%%%%%%%%
%%%%%%%%%%%%%%%%%%%%%%%%%%%%%%%%%%%%

\subsection{sXY critical line}
In the weak-link regime we can neglect the instanton--anti-instanton effects in the asymptotic flow
of the superfluid stiffness [{\it i.e.,} the term proportional to $y^{2}$ in (\ref{eq:RG_b})] in the SF phase and the critical region, and analyze the data using the simpler (\ref{eq:wk_fst_int})
with $A=f(\zeta,K(\infty))$ and
\begin{equation} \label{eq:wK_int}
w = f(\zeta,K(\infty)) - f(\zeta,K) \, .
\end{equation}
Since the $f(\zeta, x)$-function takes its maximum value at $x=1/\zeta$, see (\ref{eq:f_func}), and the value of $K$
can only decrease with the scale of distance, we immediately conclude that, if at some scale we have
$K(L) > 1/\zeta$ and simultaneously $w(L) \le f(\zeta,\zeta^{-1}) - f(\zeta,K(L))$, then the fixed point of the flow
corresponds to $w(\infty)=0$ and $K(\infty) \ge 1/\zeta$, \emph{i.e.} the phase is superfluid.
Otherwise, the flow reaches a point where $K(L') = 1/\zeta$ with $w(L')>0$ and the flow 
continues to the BG phase.

%%%%%%%%%%%%%%%%%%%%%%%%%%%%%%%
\begin{figure}[htbp]
\centering
\includegraphics[width= \linewidth]{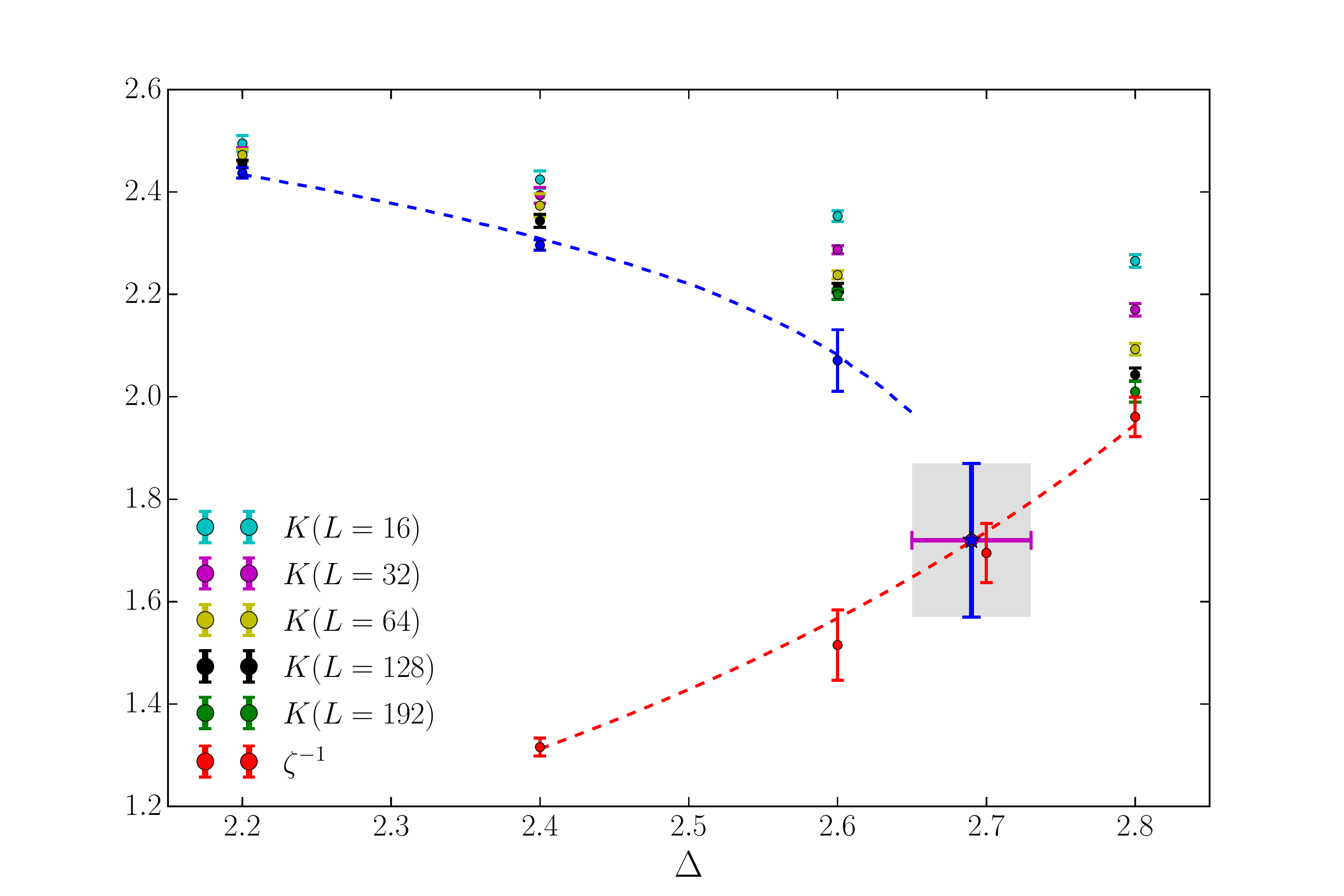}
\caption{Fine-size (see the legend) and extrapolated values (blue points and dashed line)
	of the LL parameter $K$ for different disorder strengths at $U=2.5$.
	The red dashed line (with the red dots) is the $\zeta^{-1}(\Delta)$ function obtained by linear regression of
	the data for $\zeta$ .
	The SF-BG transition point is located within the grey area where the phase is ambiguous from the fitting the RG flow due to uncertainties of $\zeta(\Delta)$ and $K(L,\Delta)$. We estimate the critical
	disorder strength to be at $\Delta_c(U)=2.69(4)$ (the magenta point and half of the horizontal width of the grey area);
	its error bar is relatively small thanks to the sharp square-root dependence of $K(\infty, \Delta )$. Correspondingly, the critical LL parameter is estimated to be $K_c(\infty)=1.72 \pm 0.15$ (the blue star and half of the vertical width of the grey area). }
\label{fig:figure2}
\end{figure}
%%%%%%%%%%%%%%%%%%%%%%%%%%%%%%%

To measure $K(L)$, we extract the compressibility $\kappa(L)$ and superfluid stiffness $\Lambda(L)$
from the particle number and winding number statistics. In the grand canonical ensemble
the probability of finding a worldline configuration with a given $N$ or $W$ number
is given by discrete Gaussian distributions
\begin{align}
W_M(M) & \propto e^{ -LTM^{2}/2 \Lambda} \;,\\
W_N(N) & \propto  e^{ -(N-\bar{N})^{2}/2 TL\kappa }  \, ,
\end{align}
where $\bar{N}$ is the average particle number. From this, the superfluid stiffness can be obtained as
\begin{equation}
\Lambda = LT \ln^{-1} \left[ \frac{W_M^{2}(0)}{W_M(1) \, W_M(-1)} \right] \, ,
\end{equation}
and compressibility as
\begin{equation}
\kappa= \frac{1}{TL} \ln^{-1} \left[  \frac{W_N^{2}([\bar{N}])}{ W_N([\bar{N}]+1) \, W_N([\bar{N}]-1)} \right] \, ,
\end{equation}
where $[\bar{N}]$ is the closest integer to $\bar{N}$.

The protocol of determining the critical point is as follows.
We fix the value of $U$ and start with measuring the $\zeta (\Delta)$ dependence
(all data points can be perfectly fit to a linear dependence).
Next, we compute $K(L,\Delta)$ values for a number of different system sizes and $\Delta$-points.
$K(L,\Delta)$ is reported as the median of the distribution over several hundred (up to a thousand)
disorder realizations.
Finally, for each value of $\Delta$, we employ (\ref{eq:RG_wk_1}) and (\ref{eq:RG_wk_2}) to extract the $w(L)$-function
by fitting finite-size data to the flow equations. Depending on the result, we then either derive
the thermodynamic limit answer for $K(\infty)$ from (\ref{eq:wK_int})
or conclude that the flow is to the BG phase.
To improve our estimate of the critical disorder strength, given a finite mesh in $\Delta$,
we interpolate $K(L,\Delta)$ data between the points using linear fits and proceed with the flow analysis
as described above; higher-order polynomial fits produce similar results within the error bars.
The critical parameter $\Delta_c(U)$ is then found from the intersection of $K(\infty , \Delta)$
and $\zeta^{-1} (\Delta )$ curves, see figure~\ref{fig:figure2}. Its error bar is mostly determined
by the uncertainty on the closest $\zeta$ and $K(\infty)$ points.

%%%%%%%%%%%%%%%%%%%%%%%%%%%%%%%
\begin{figure}[htbp]
\centering
\includegraphics[width=\linewidth]{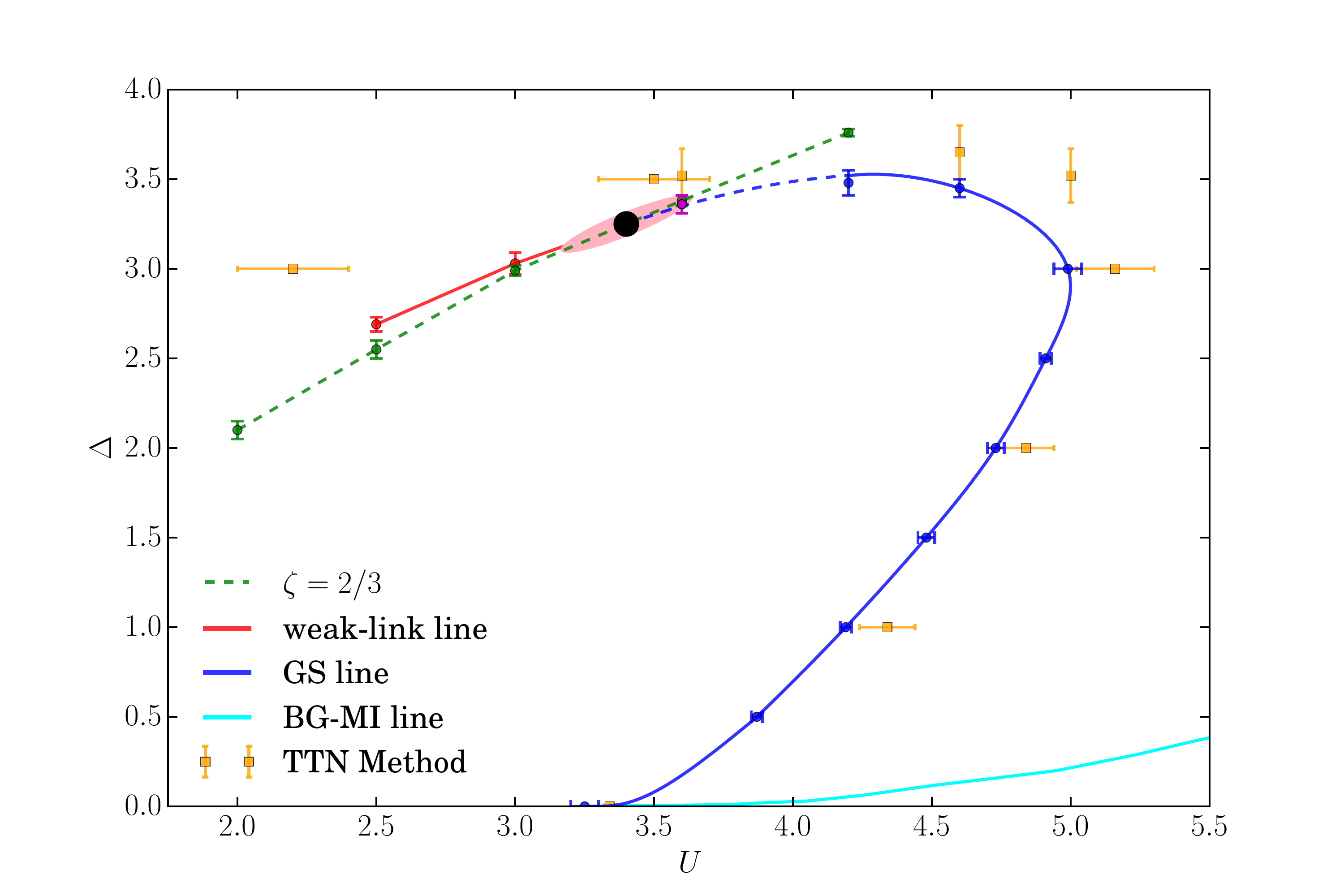}
\caption{Ground-state phase diagram of the 1D disordered Bose-Hubbard model at unit filling factor.
	The sXY and GS critical lines are shown in red and blue, respectively.
	The intersection of the interpolated $\zeta=2/3$ line (dashed green) with any of the other critical lines
	determines the tri-critical point (black dot within the pink uncertainty region).
	The cyan line shows the gaps of the Mott insulator in the disorder-free system taken from \cite{Kuehner}, which signals the transition between the Mott insulator and the BG phase in the presence of disorder.
	We also show the $K(\infty)=3/2$ line obtained by the Tree Tensor Network (TTN) method \cite{TTN} (orange), which
	agrees with our GS-line within the error bars. As expected, in the weak-link regime, the TTN line ends inside the BG phase.}
\label{fig:figure3}
\end{figure}
%%%%%%%%%%%%%%%%%%%%%%%%%%%%%%%

\subsection{Tri-critical point and the Giamarchi--Schulz criticality}
The tri-critical point $(U_{*}, \Delta_{*})$ separating the sXY and GS universality classes can be found
from the intersection of the sXY critical and $\zeta=2/3$ lines. Two circumstances help us to locate it relatively accurately. On the one hand, from the analysis performed in section \ref{sec:IV} we conclude that
the sXY critical line can be smoothly interpolated all the way to the intersection point. On the other hand,
the sXY critical point $\Delta_C(U=3.6)=3.36$ [the black dot in figure~\ref{fig:figure3}]
deduced by the protocol described in the previous subsection, landed on the $\zeta=2/3$ line [located at
$(U=3.6,\, \Delta=3.38)$] within error bars. This basically eliminates the need for
determining the tri-critical point from the intersection of interpolated curves. The procedure predicts
\begin{equation}
U_{*} = 3.40 \pm 0.23,  \qquad \Delta_{*} = 3.25 \pm 0.15 \, .
\end{equation}
The value for $U_{*}$ is remarkably close to the critical value for the superfluid to Mott insulator transition in the absence of disorder.

By knowing the slopes of the sXY critical and $\zeta=2/3$ lines in the $(\Delta, U)$-plane and the location of the tri-critical point,
we are in a position to relate the integration constant $D$ and the small parameter $\delta$, controlling the shape of
the phase diagram in the vicinity of $(U_{*}, \Delta_{*})$, to the Hamiltonian parameters.
Using a linear expansion about the $D=0,\, \delta=0$ point
\begin{align}
    D      &= A_{11} (U-U_{*}) - A_{12} (\Delta - \Delta_{*})  \, , \label{DA} \\
    \delta &= A_{21} (U-U_{*}) - A_{22} (\Delta - \Delta_{*})  \, ,
\label{deltaA}
\end{align}
and numerical data determining the $D=0$ and $\delta=0$ curves, we find that
\begin{equation}
A_{11}/A_{12} = 0.55 \pm 0.11 \, , \qquad  A_{21}/A_{22} = 0.65 \pm 0.06 \, .
\label{A11A21}
\end{equation}

Next, the dependence of $K(\infty)$ on $\Delta$ in the SF phase in the weak-link regime
allows us to obtain $A_{12}$. Specifically, near the tri-critical point (\ref{eq:xw_eta}) implies
\begin{equation}
K(\infty , \Delta ) - 1/\zeta  = \sqrt{D(\Delta)} \, ,
\end{equation}
This analysis allows us to determine the derivative of $D$ with respect to $\Delta$ and results in
\begin{equation}
A_{12}  = 1.0 \pm 0.2 \, .
\label{A12}
\end{equation}
Similarly, $A_{22}$ controls the slope of the $\zeta (\Delta)$ line at fixed $U$.
From the data sets computed at $U=3.0$ and $U=3.6$ and linearly extrapolated to the tri-critical point,
we find
\begin{equation}
A_{22} = 0.56 \pm 0.14  \, .
\label{A22}
\end{equation}

%%%%%%%%%%%%%%%%%%%%%%%%%%%%%%%%%%
\begin{figure}[htbp] 
	\centering
	\includegraphics[width=\linewidth]{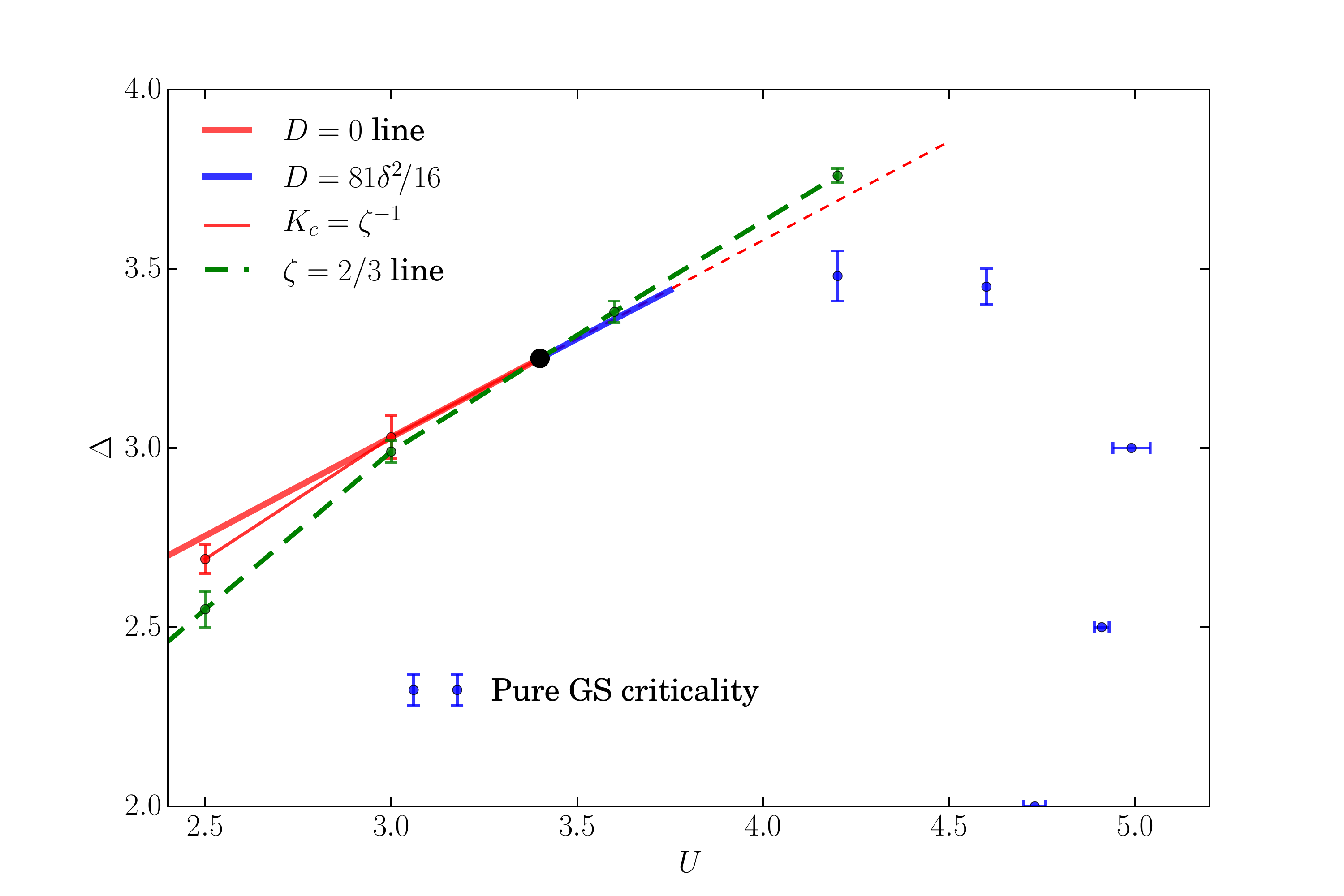}
	\caption{The SF-BG phase diagram in the vicinity of the tri-critical point.
		(\ref{DA}), (\ref{deltaA}), (\ref{A11A21}), (\ref{A12}), and (\ref{A22})
		were used to plot the $D=0$ (bold red and dashed red), $\delta=0$ (dashed green), and $D=(81/16) \delta^{2}$ (bold blue)
		predictions for sXY, $\zeta=2/3$, and GS lines, respectively. We also plot
		our numerical data for the same lines (same color scheme as in figure~\ref{fig:figure3}).
	}
	 \label{fig:figure4}
\end{figure}
%%%%%%%%%%%%%%%%%%%%%%%%%%%%%%%%%%%%%

We are all set to make a quantitative prediction for the structure of the phase diagram
in close vicinity of the tri-critical point, including the location of the GS-line
(the protocol of calculating the GS line away from the tri-critical point is outlined below).
The result is shown in figure \ref{fig:figure4} where (\ref{DA}) and (\ref{deltaA})
are used to plot the $D=0$, $\delta=0$, and $D=(81/16) \delta^{2}$ [see (\ref{eq:GS_c})]
curves for sXY, $\zeta=2/3$, and GS lines, respectively.
By observing reasonable agreement between this prediction and an independent calculation
of the GS-line relatively far from the tri-critical point we validate the proposed theory.

Deep in the GS regime, when the weak link term $w$ in the RG equations can be neglected,
we are back to the standard $XY$-universality class analysis. From
\begin{align}
    {\frac{dy}{d \ell} } &= { (3/2-K)y}  \, , \label{eq:RG_GS_y} \\
    {\frac{dK}{d \ell} } &= { -y^{2} K^{2} }\, , \label{eq:RG_GS_K}
\end{align}
we readily obtain the first integral as
\begin{equation} \label{eq:GS_Intg}
2 \ln K + 3/K = y^{2} + G \, ,
\end{equation}
where $G$ is the integration constant. The thermodynamic state can be established by solving the RG flow just as we did for the weak-link regime. In the SF phase, the flow has a fixed point at $y=0$ and $K(\infty) \ge 3/2$. Otherwise, if the finite-size system can reach
a state with $K(L)=3/2$ and $y>0$, the flow will continue towards the BG phase with $K(\infty)=0$. Numerically,
we first fit the finite-size data to the RG flow to determine $G$ and then use the above-mentioned property of the
first integral (\ref{eq:GS_Intg}) to determine the phase. Technically, this protocol is nearly identical
to the one used in the weak-link regime and we do not repeat it here. The resulting GS-line is shown in figure~\ref{fig:figure3}.

%%%%%%%%%%%%%%%%%%%%%%%%%%%%%%%%%%%%%%%%%%%%%%%%%%%%%%%%%%%%%%%%%%%%%
\section{Concluding remarks}
\label{sec:VI}
%%%%%%%%%%%%%%%%%%%%%%%%%%%%%%%%%%%%%%%%%%%%%%%%%%%%%%%%%%%%%%%%%%%%%

The importance of our successful application of the sXY criticality theory
to the SF-BG transition in model (\ref{BH}) is two-fold.
First, we have corroborated the analytic theory of the interplay between the two university classes
and the structure of the phase diagram in the vicinity of the tri-critical point.
Second, we establish the qualitative and quantitative  behavior of the
ground-state phase diagram (and finite-size properties) of the Hamiltonian (\ref{BH}) numerically.
The theory rests on a rather non-trivial postulate of the existence of an irrenormalizable
power-law distribution of microscopic weak links.
While being self-consistent---and, in this sense, rendering the theory asymptotically exact---the
postulate can hardly be proven as a theorem. Therefore, the data in figure~\ref{fig:figure1}
demonstrating excellent agreement with the $\zeta$-postulate is
at least as important as the phase diagram shown in figure~\ref{fig:figure3}.

The ground-state phase diagram of (\ref{BH}) in the $(U, \Delta)$ plane features a characteristic line defined by
the condition $\zeta(U, \Delta) = 2/3$. Strictly speaking, this line is well defined only in the superfluid phase
and at the SF-BG phase boundary. However, the exponential divergence of the correlation length on approach
to the SF-BG critical point guarantees that the $\zeta = 2/3$ line remains meaningful even
inside the BG phase [see figure~\ref{fig:figure3}]; the data presented in figure~\ref{fig:figure1} further illustrate this point.
On the  $\zeta < 2/3$ side from the $\zeta = 2/3$ line, the sXY criticality preempts the GS scenario.
This, in particular, means that superfluidity with $ 3/2 < K(L) \leq 1/\zeta$ in this part of the phase diagram is
guaranteed to be a finite-size effect, since the $L \to \infty$ phase is BG.

At the intersection of the  $\zeta = 2/3$ line with the SF-BG  phase boundary
there is a tri-critical point separating the GS and sXY criticalities.
According to our analysis, the phase boundary remains smooth at the tri-critical point but the curvature is discontinuous.
For purely numeric reasons the angle between the $\zeta = 2/3$ line and the phase boundary happens to be rather small.
As a result, despite accurately identifying the positions of both the $\zeta = 2/3$ line and the phase boundary,
the uncertainty in the location of the tri-critical point remains relatively large.
Another consequence of the small angle intersection is that the critical value of $K$ on the sXY line
is only slightly higher than the GS value of $3/2$. Under such circumstances, a brute-force observation
of the violation of the GS scenario in the vicinity of the tri-critical point becomes problematic (cf. \cite{TTN}) even though our data in figure~\ref{fig:figure2} are not compatible with GS even when done with a brute force analysis.

Tracing the fate of the sXY line far away from the tri-critical point (in the region of small $U$ and $\Delta$)
requires a substantial numerical effort and goes beyond the scope of this paper.

\acknowledgments{
We acknowledge support from the National Science Foundation under Grant PHY-1314735, MURI Program ``New Quantum Phases of Matter" from AFOSR, FP7/ERC starting grant No.\ 306897 and FP7/Marie-Curie CIG grant  No.\ 321918. Part of the simulation is done in the MGHPCC.
}
%

%%%%%%%%%%%%%%%%%%%%%%%%%%%%%%%%%%%%%%%%%%%%%%%%%%%%%%%%%%%%%

\end{document}